\documentclass[%
 reprint,
 amsmath,amssymb,
 aps,nofootinbib,longbibliography   
]{revtex4-1}

\usepackage{amsmath}	
\usepackage{amsthm}		
\usepackage{amssymb}	
\usepackage{eufrak}
\usepackage{datetime}
\usepackage{graphicx}
\usepackage{color}
\usepackage{verbatim}
\usepackage{bm}
\usepackage{braket}
\usepackage{natbib}
\usepackage{cancel}
\usepackage{xcolor}
\usepackage{slashed}
\usepackage{braket}
\usepackage{caption}
\usepackage{subcaption}
\usepackage{amsfonts} 
\usepackage{float}
\smallskipamount
\usepackage{feynmp}
\usepackage{tikz-feynman}
\tikzfeynmanset{compat=1.1.0}

\usepackage[colorlinks=true, citecolor=midblue, linkcolor=midblue, urlcolor=midblue]{hyperref}

\usepackage{setspace}

\settimeformat{ampmtime}

\definecolor{grey}{rgb}{0.4,0.4,0.4}
\definecolor{dullmagenta}{rgb}{0.4,0,0.4}
\definecolor{darkblue}{rgb}{0,0,0.4}
\definecolor{midblue}{rgb}{0,0,0.5}
\definecolor{midred}{rgb}{0.5,0,0}
\definecolor{orange}{rgb}{1,0.5,0}
\definecolor{lightbrown}{rgb}{0.75,0.5,0.25}
\definecolor{tan}{cmyk}{0.14,0.42,0.56,0}
\definecolor{djunglegreen}{cmyk}{0.99,0,0.52,0}
\definecolor{lightgreen}{rgb}{0,1,0}
\definecolor{olivegreen}{cmyk}{0.64,0,0.95,0.40}
\definecolor{midgreen}{rgb}{0.0,0.675,0.0}
\definecolor{darkgreen}{rgb}{0,0.5,0}

\settimeformat{ampmtime}

\begin{document}
\title{Dynamics of Confined Monopoles \\
  and Similarities with Confined Quarks}

\author{Gia Dvali}
\affiliation{
	Arnold Sommerfeld Center,
	Ludwig-Maximilians-Universit{\"a}t,
	Theresienstra{\ss}e 37,
	80333 M{\"u}nchen,
	Germany}
\affiliation{
	Max-Planck-Institut f{\"u}r Physik,
	F{\"o}hringer Ring 6,
	80805 M{\"u}nchen,
	Germany}
	
\author{Juan Valbuena-Bermudez}
\email{juanv@mpp.mpg.de}

 \affiliation{
 	Arnold Sommerfeld Center,
 	Ludwig-Maximilians-Universit{\"a}t,
 	Theresienstra{\ss}e 37,
 	80333 M{\"u}nchen,
 	Germany}
 \affiliation{
 	Max-Planck-Institut f{\"u}r Physik,
 	F{\"o}hringer Ring 6,
 	80805 M{\"u}nchen,
	Germany}

 \author{Michael Zantedeschi}
 \email{michaelz@mpp.mpg.de}
\affiliation{
 	Arnold Sommerfeld Center,
 	Ludwig-Maximilians-Universit{\"a}t,
	Theresienstra{\ss}e 37,
 	80333 M{\"u}nchen,
 	Germany}
 \affiliation{
 	Max-Planck-Institut f{\"u}r Physik,
 	F{\"o}hringer Ring 6,
 	80805 M{\"u}nchen,
 	Germany}

\date{\formatdate{\day}{\month}{\year}, \currenttime}

\begin{abstract}
In this work, we study the annihilation of a pair of `t Hooft-Polyakov monopoles due to confinement by a string. We analyze the regime in which the scales of monopoles and strings are comparable.
We compute the spectrum of the emitted gravitational waves and find it to agree with the previously calculated point-like case for wavelengths longer than the system width and before the collision. However, we observe that in a head-on collision, monopoles are never re-created. Correspondingly, not even once the string oscillates.
Instead, the system decays into waves of Higgs and gauge fields.   
We explain this phenomenon by the loss of coherence in the annihilation process. Due to this, the entropy suppression 
makes the recreation of a monopole pair highly
improbable. We argue that in a similar regime, 
analogous behaviour is expected for the heavy quarks connected by a QCD string. There too, instead of re-stretching a long string after the first collapse, the system hadronizes and decays in a high multiplicity of mesons and glueballs. We discuss the implications of our results.   
\end{abstract}
\maketitle
\section{Introduction}
It is well known~\cite{langacker1980magnetic,Lazarides:1981fv,Vilenkin:1982hm,Vilenkin:2000jqa} 
that monopoles carrying the opposite magnetic charges  under a $U(1)$ gauge symmetry  become connected 
by a string when $U(1)$ is Higgsed. The string 
represents a magnetic flux tube of Nielsen-Olesen type~\cite{Nielsen:1973cs}.  
As long as the mass of the monopole is larger than the  
scale of the string tension, energy per unit length $\mu$, the breakup of the string via nucleation 
of monopole pairs is exponentially unlikely~\cite{Vilenkin:1982hm}.  
 
This system shares some similarity with a 
quark-anti-quark pair connected  by a QCD string in a confining gauge theory. The QCD string represents a flux tube of the color electric field. The string tension, $\mu = \Lambda^2$, is set by the QCD scale, $\Lambda$. 
As long as all quarks in the theory are heavier than $\Lambda$, 
the probability of breaking the string by a pair creation is exponentially small. 

Due to this analogy, studying monopoles connected by a magnetic string 
can serve as a useful test-laboratory for understanding certain features of confined heavy quarks.  

The above systems have a number of interesting applications
in particle physics and cosmology. 

For example, recently a novel mechanism for producing the primordial black holes was proposed in~\cite{Dvali:2021byy}. Therein, quark pairs, produced and diluted in the inflationary era, are confined in the late Universe.  Upon horizon re-entry, they collapse and  form black holes (BHs) due to the large amount of energy stored in the flux tubes connecting them. Given the constant acceleration of quarks sourced by the string, gravitational waves (GWs) of frequency comparable to the inverse of the horizon size are produced. 
Analogous considerations could be applied to the case of confined monopoles~\cite{Matsuda:2005ey}.

Previous calculations of the radiated GW spectrum 
were performed by Martin and Vilenkin~\cite{Martin:1996cp}
in the point-like approximation, in which the size of monopoles as well as the width of the string are set to zero.  
In this limit, they obtained  the following emitted power for a large range of frequencies 
\begin{equation}
    P_{\rm n} \sim  \frac{\Lambda^4}{M_{\rm p}^2}\frac{1}{n},
\end{equation} 
$n$ being the frequency number, $M_{\rm p}$ the planck mass, and $\Lambda$ the confining scale. This relation was derived considering a pair of monopoles connected by a string and is expected to be valid in the case of confined quarks too. Such sources could explain the recent hints of stochastic GW background obtained from pulsar timing arrays~\cite{NANOGrav:2020bcs,Antoniadis:2022pcn}. Moreover, given the flatness of the resulting energy density across several orders of frequency~\cite{Leblond:2009fq}, in the future, it will be possible to cross-check with other gravitational wave detectors sensitive to the lengths shorter than the pulsar timing arrays. 

Another scenario for which our study is relevant is the Langacker-Pi mechanism~\cite{langacker1980magnetic}. In an attempt to solve the monopole abundance problem, the theory ensures  a temporary (thermal) window in which the $U(1)$ group associated to the monopole charge is broken, leading to their confinement.

At lower temperatures the $U(1)$ symmetry is restored again. 
This mechanism can be achieved by adequately choosing the spectrum and couplings of the theory~\cite{weinberg1974gauge}.
The system therefore has a finite window of opportunity for getting rid of monopoles. If monopoles connected by string can oscillate for too long, this window of opportunity is insufficient for solving the monopole problem.  

The goal of the present work is to analyze the dynamics of a monopole/anti-monopole pair in the confined phase in detail. In order to do so, we considered a $SU(2)$ gauge theory and chose a simple scalar sector capable of achieving the above-mentioned configuration via spontaneous symmetry breaking: a scalar field in the adjoint representation, and a complex scalar doublet. The former breaks (Higgses) the gauge group $SU(2)$ to $U(1)$, therefore admitting t'Hooft-Polyakov monopoles as a solution~\cite{Hooft1974MagneticTheories,Polyakov1975CompactCatastrophe}. The latter breaks the residual $U(1)$ gauge group leading to the confinement of the associated ``magnetic flux''. Our study covers the regime in which the monopole size and the string width are comparable. The similarities with confined quarks are established in the analogous regime.
 
It turns out that the point-like limit  
 approximates very well the part of the classical dynamics 
 in which the monopole separation is much larger then 
the characteristic width of the system.   
However, beyond this regime we observe some new features. 
 
Naively,  it is expected that a collapsing straight string 
performs several oscillations.   
That is, one would think that after shrinking, the end points
(monopoles) scatter and fly apart stretching a long string again.  
In this way, the string would contract and expand with certain periodicity, as some sort of a rubber band. 
   
However, we observe that in head-on collision 
the outcome is very different.  After the first shrinkage the string never recovers. 
Instead, the entire 
energy is converted into the waves of Higgs and gauge particles.  
These waves can also be thought of as large number of overlapping short strings. 

We explain this phenomenon and argue that in analogous kinematic regime the similar effect takes place
in case of confined quarks. 
In this particular regime, in both cases, the outcome can be 
understood as the result of the entropy suppression for
production of a highly coherent state in a collision process~\cite{Dvali:2020wqi}. 
Due to this, instead of stretching a long string,  the system prefers to produce many particles (short strings) which have a much higher entropy.  
In case of QCD, the collapse of a long string results into a high multiplicity of glueballs  (closed strings) and mesons (open strings). 

 We also point out that inability of monopole and 
antimonopole to going through each other, 
  falls in the same category as the suppression 
of the passage of a magnetic monopole through a domain wall,
studied in~\cite{Dvali:1997sa}. 
In that example, the domain wall provides a support base for unwinding the monopole, similar to the role of the antimonopole in the present case. The recreation of the monopole state on the other side of the wall is unlikely due to the insufficiency of the microstate entropy of the monopole for overcoming the exponential suppression  of the corresponding multi-particle amplitude~\cite{Dvali:2020wqi}.
This leads to the ``erasure" of monopoles by domain walls.  
In~\cite{Dvali:1997sa}, this mechanism was used to solve the cosmological monopole problem in grand unified theories. 
However, the phenomenon of erasure is of broader fundamental interest. In particular, this is indicated by the similarities
between the erasure processes of confined quarks and confined monopoles discussed in the present paper.

It emerges that in the studied regime, the processes 
of the collapse of the confined pairs in both theories 
are governed by the same universal effect:  the 
exponential suppression of production 
of a high occupation number (coherent) state, albeit of insufficient entropy~\cite{Dvali:2020wqi}   

The GW spectrum produced by confined monopoles 
is appropriately captured by the point-like result for scales larger than the monopole width. As expected, we observe non-negligible corrections to the power spectrum for scales comparable to the monopole radius, where the emitted radiation is boosted, therefore providing corrections to the GWs emission produced by the confinement dynamics. 

We expect that our results have implications for the collapse of the generic bounded strings such as the string-theoretic strings bounded by $D$-branes~\cite{Copeland:2003bj, Dvali:2003zj}.
 
The paper is organized as follows. 
First we discuss the system of confined monopoles and 
study it numerically.  Next, we explain the underlying 
physics that is shared by confined quarks and monopoles.    
We then study emission of gravitational waves. 
Finally we discuss sphalerons and give outlook and conclusions. 
 
 
\section{Setup}
We will work with a $\text{SU}(2)$ gauged field theory that contains a scalar field in the adjoint representation, $\varphi^a$ (a = 1, 2, 3), a scalar field in the fundamental representation, $\psi$, and gauge fields, $W^a_\mu$. The Lagrangian of the system is given by
\begin{equation}
    \mathcal{L}=\frac{1}{2}D_\mu \varphi^a D^\mu \varphi^a +
    (D_\mu \psi)^\dag D_\mu \psi
    -\frac{1}{4}{W^a}_{\mu\nu}{W^a}^{\mu\nu}-
    V(\varphi,\psi)\label{eq:Lagrangian}
\end{equation}
where summation over repeated $\text{SU}(2)$ indices is understood, and the field strengths for the gauge field is 
\begin{equation}W_{\mu\nu}^a=\partial_\mu W^a_\mu-\partial_\nu W^a_\mu + g \epsilon^{abc}W^b_{\mu} W^c_{\nu}.\end{equation}
The covariant derivatives are defined as
  \begin{align}
 D_\mu \varphi^a &= \partial_\mu \varphi^a + g \epsilon^{abc} W^b_\mu \varphi^c,\\
 D_\mu \psi &= \partial_\mu \psi- i g \frac{\sigma^a}{2} W^a_\mu \psi,
\end{align}
and the potential is given by~\cite{Kibble:2015twa}
\begin{equation}
\label{eq:potential}
    V(\varphi,\psi)=
    \frac{\lambda}{4}(\varphi^a \varphi^a -\eta^2)^2 
    +\frac{ \tilde \lambda}{2} (\psi^\dag \psi - v^2)^2 
    + c\, \psi^\dag \sigma^a \psi \varphi^a .
\end{equation}
As the first stage of symmetry breaking, we give vacuum expectation value to the adjoint field while keeping $\psi=0$. The system admits 't Hooft-Polyakov monopoles~\cite{tHooft:1977nqb,Polyakov1975CompactCatastrophe}. As $\psi$ acquires vacuum expectation value, the $SU(2)$ gauge symmetry is Higgsed to $U(1)$, and the magnetic flux of a monopole is trapped into a tube which can end on an anti-monopole. In this way, monopoles become confined. The dynamics of such configuration are the main focus of this work. 
\begin{figure}[t]
    \centering
    \includegraphics[width=0.3\textwidth]{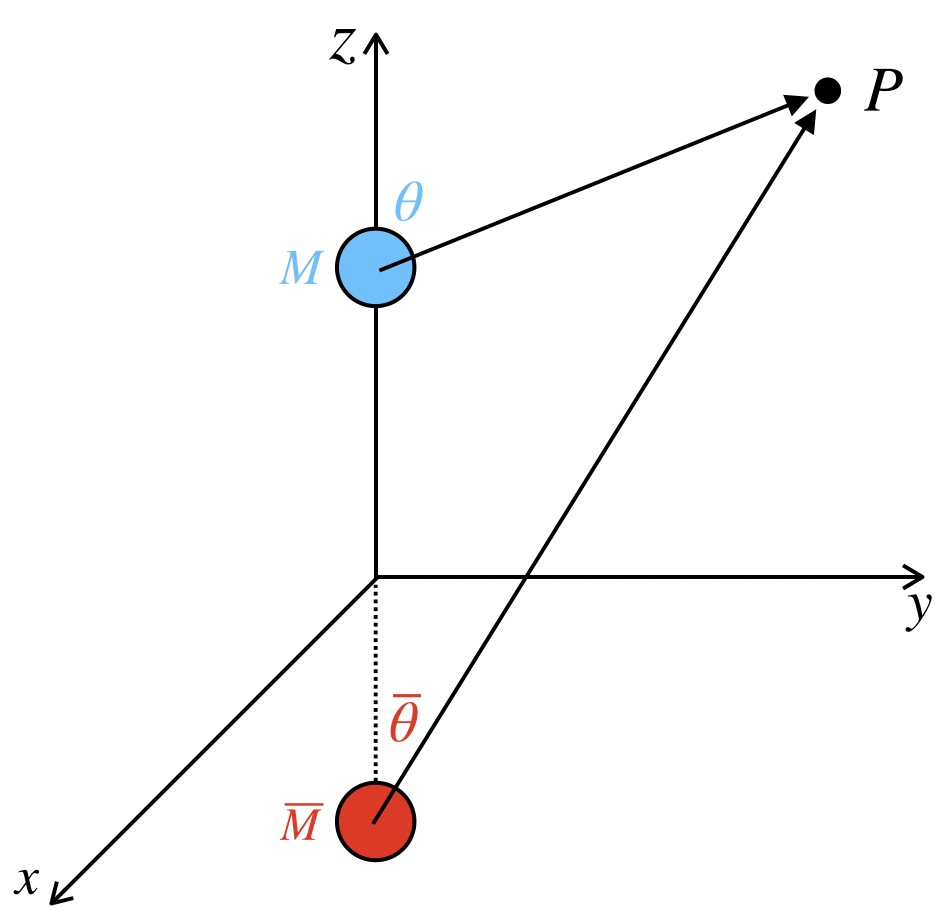}
    \caption{A sketch of the initial monopole/anti-monopole initial configuration. }
    \label{fig:sketchinitial}
\end{figure}
The initial configuration utilized in this work is sketched in Fig.~\ref{fig:sketchinitial}. The monopoles are aligned along the $z-$axis at a distance $d$. In the figure $\theta$ and $\overline \theta$ denote the respective monopole and antimonopole position azimuthal angle. In the approximation when the distance $d$ is much longer than the monopole size the string configuration can be derived as follows.  
For $\overline\theta =0$, a monopole should be recovered $\psi \propto (\cos{\theta/2}, \sin{\theta/2}e^{i\phi})^t$~\cite{Nambu:1974zg,Nambu1977String-likeTheory}, while for $\theta=\pi$ an antimonopole should be obtained $\psi\propto (\sin\overline\theta/2, \cos \overline \theta/2 e^{i \phi})^t$, $\phi$ being the polar angle. Therefore the string configuration is given by \cite{Vachaspati:1994ng, Vachaspati2016Monopole-antimonopoleScattering,Saurabh2017Monopole-antimonopolePotential}
\begin{equation}
 \psi\propto \begin{pmatrix}
  \sin (\theta/2) \sin (\Bar{\theta}/2)e^{i\gamma} + \cos(\theta/2)\cos(\Bar{\theta}/2)\\ 
  \sin (\theta/2) \cos (\Bar{\theta}/2)e^{i\phi} - \cos(\theta/2)\sin(\Bar{\theta}/2)e^{i(\phi-\gamma)}
\end{pmatrix}
\label{eq:String-Ansatz2}    
\end{equation}
with $\gamma$ accounting for the possibility of twisting the antimonopole w.r.t. the monopole. In fact, for $\theta=\pi$, $\psi$ corresponds to the above mentioned antimonopole under the shift $\phi \rightarrow \phi +\gamma$. Above the configuration, for $\theta=\overline{\theta}=0$, $\psi= (v,0)^t$, while below $\theta=\overline{\theta}=\pi$, $\psi = (v e^{i\gamma},0)^t$. Finally between the two monopoles, $\theta=\pi$ and $\overline{\theta}=0$, $\psi\propto (e^{i\phi},0)^t$ corresponding to the unit winding string. 

Asymptotically the string is proportional to the positive eigenvector of the third Pauli matrix. Since we choose $c<0$ in the last term of the potential \eqref{eq:potential}, the adjoint field direction $\hat{\varphi}^a$ of the monopoles can be built asymptotically as \cite{Nambu:1974zg,Nambu1977String-likeTheory}
\begin{equation}
\label{eq:adjdirection}
    \hat \varphi ^a = \frac{1}{v^2}\psi^\dagger \tau^a \psi, \quad a=1,2,3
\end{equation}
where $\tau^a$ denotes the three Pauli matrices (see Fig.~\ref{fig:examplephi}). 

In the next section we analyze the monopole/anti-monopole configuration after the first phase transition, ignoring the doublet. Although this was already explored by Vachaspati and Saurabh~\cite{Vachaspati2016Monopole-antimonopoleScattering, Saurabh2017Monopole-antimonopolePotential}, it serves as a useful exercise before turning to the symmetry broken phase.

\section{Monopole/anti-monopole system}
The explicit equations of motion can be found in~Appendix~\ref{app:Fields_eq}. From now, we work in energy units of $\eta^{-1}$ and set the gauge coupling $g=1$.
Thus $\lambda$ is the parameter in the theory that controls the mass and size of the monopoles.

For a (spherically symmetric) monopole field configurations, we use the following ansatz
\begin{equation}
\label{eq:TP-Ansatz1}
	\varphi^a =h(r) \hat{r}^a, \quad W^a_i = \frac{(1-k(r))}{r} \epsilon^{aij}\hat{r}^j
\end{equation}
where $r$ is the radial coordinate and $\hat{r}^a=r^a/|\Vec{r}|$. Under ansatz~\eqref{eq:TP-Ansatz1}, the equations of motion become:
\begin{align}
    h''(r) +\frac{2}{r}h'(r) &= \frac{2}{r^2}k(r)^2h(r) - \lambda(h(r)^2-1)h(r),\\
    k''(r) &= \frac{1}{r^2}(k(r)^2 - 1)k(r) + h(r)^2k(r),
\end{align}
with asymptotic conditions
 \begin{align}
      h(r)\xrightarrow{r\rightarrow 0} 0, \ & \ k(r)\xrightarrow{r\rightarrow 0}1,\\
     h(r)\xrightarrow{r\rightarrow\infty}1, \ & \ k(r)\xrightarrow{r\rightarrow\infty}0.
 \end{align}
The above equations were solved numerically in order to obtain the monopoles profile.\\

Finally the ansatz for the initial adjoint field configuration is given by
\begin{equation}
    \varphi^a = h(r_m)h(\Bar{r}_m)\hat{\varphi}^a,
    \label{eq:MMbar-Ansatz-varphi}
\end{equation}
with $r_m$ ($\Bar{r}_m$) denoting the monopole (anti-monopole) coordinate center and $\hat{\varphi}^a$ is defined in \eqref{eq:adjdirection}.
\begin{figure}[t]
    \centering
    \includegraphics[width=0.3\textwidth]{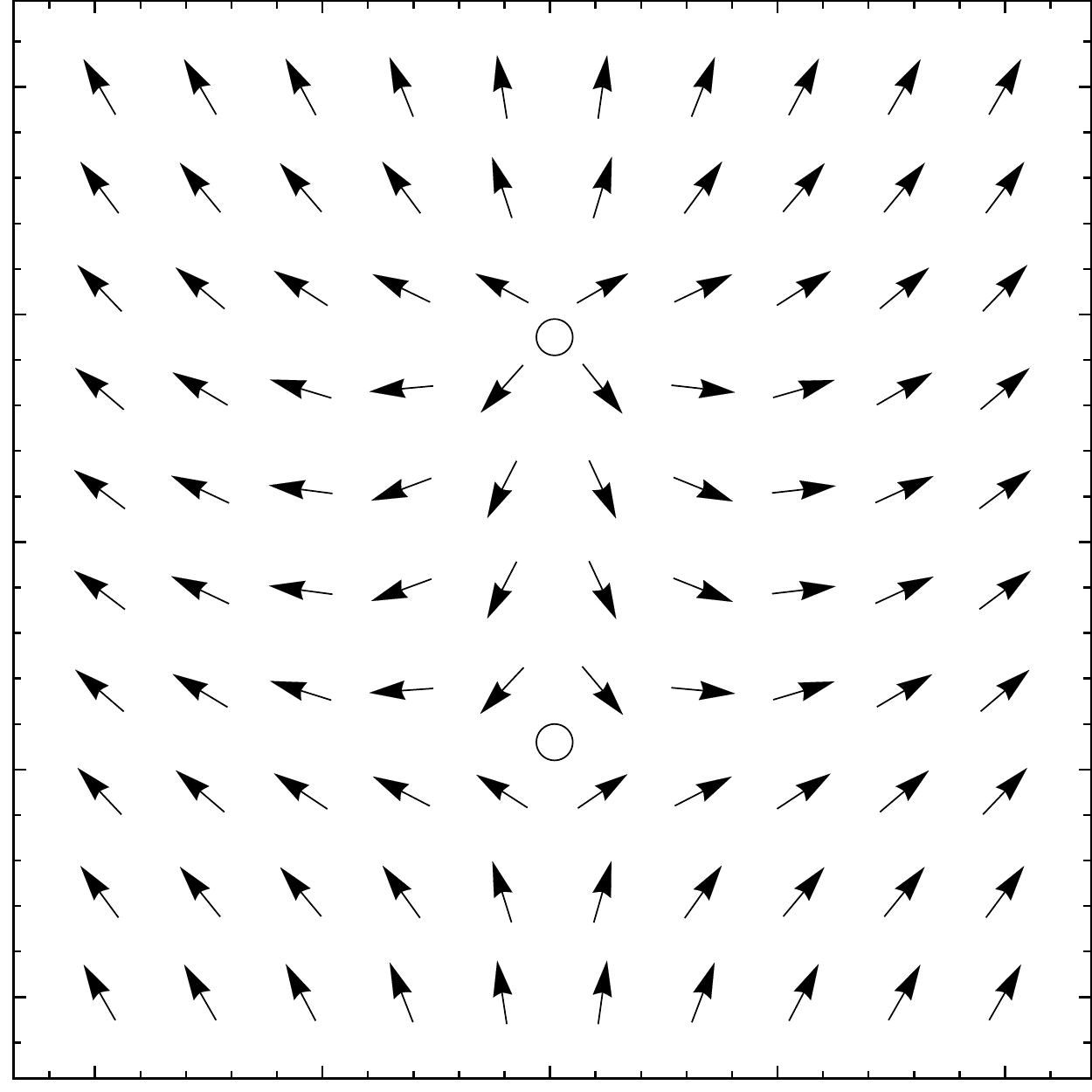}
    \caption{Example of field configuration $\hat\varphi^a$ for twist $\gamma=0$ according to Eq.~\eqref{eq:MMbar-Ansatz-varphi} in the $yz$ plane. The circles denote the position of the monopole and antimonopole. }
    \label{fig:examplephi}
\end{figure}
An example of such configuration is shown in Fig.~\ref{fig:examplephi}.

The stationary ansatz for the gauge fields considered by Vachaspati and Saurabh~\cite{Vachaspati2016Monopole-antimonopoleScattering, Saurabh2017Monopole-antimonopolePotential}, follows from the requirement that the covariant derivative of the Higgs isovector vanish  at spatial infinity, $D_\mu \hat{\varphi}|_{r\rightarrow\infty} = 0$. This gives
\begin{equation}
W^a_\mu= -(1-k(r_m))(1-k(\Bar{r}_m))\epsilon^{abc}\hat{\varphi}^b \partial_\mu \hat{\varphi}^c .
    \label{eq:MMbar-Ansatz-W}
\end{equation}
As expected, and verified in \cite{Saurabh2017Monopole-antimonopolePotential}, the monopole/anti-monopole are attracted to each other due to a ``magnetic'' Coulomb-like interaction (for distances much bigger than the monopole size). Moreover, the potential energy is also affected by the initial system twist parametrized by $\gamma$ \cite{taubes1982existence} - such a correction, however, is exponentially suppressed at large distances. The verification of these properties served as a valuable check of the numerics presented in this work.

\section{Monopole/Anti-monopole connected by strings}
We solve field equations \eqref{eq:EOM-Varphi}-\eqref{eq:EOM-Gamma}, with an initial configuration of the  monopole/anti-monopole given by \eqref{eq:MMbar-Ansatz-varphi} and \eqref{eq:MMbar-Ansatz-W}. Before starting the dynamical evolution a numerical relaxation of the configuration was performed (c.f. \cite{Saurabh2017Monopole-antimonopolePotential}).

Asymptotically $\varphi^a= \eta \delta^{a3}$, implying that $\psi$, due to the last interaction term in \eqref{eq:potential}, needs to be proportional to the positive $\sigma^3$ eigenvector $\langle|\psi|\rangle\propto (1,0)^{t}$, as correctly implied by ansatz \eqref{eq:String-Ansatz2}\footnote{we used $c<0$ in our numerical simulations}.

Above the pair, $\psi$ ansatz minimizes the interaction energy. As it crosses the monopole, the configuration becomes singular at the south pole and its phase is flipped by $2 \pi$. It follows that between the two monopoles, along the axial axes, $\psi \propto (0,e^{i \phi})^t$, $\phi$ being the polar angle. By construction the associated winding number is one. Finally, as it reaches the anti-monopole, the string flips by $2\pi$ again and it becomes proportional to $\psi \propto (e^{i \gamma},0)^t$.
While minimization of scalar potential is independent on the value of $\gamma$, the same is not true for both the kinetic terms and the gauge sector. In fact, $\gamma\neq 0$ generates a repulsive interaction  \cite{Saurabh2017Monopole-antimonopolePotential} between the quark/anti-quark pair which can be easily shown to be maximal for $\gamma = \pi$. Unless otherwise stated, from now on we will focus on the case $\gamma =0$.

As the ``magnetic'' field of the pair is confined into tubes, the monopoles accelerate towards each other, turning rapidly, relativistic and annihilate.
\begin{figure}[h]
     \centering
         \includegraphics[width=.45\textwidth]{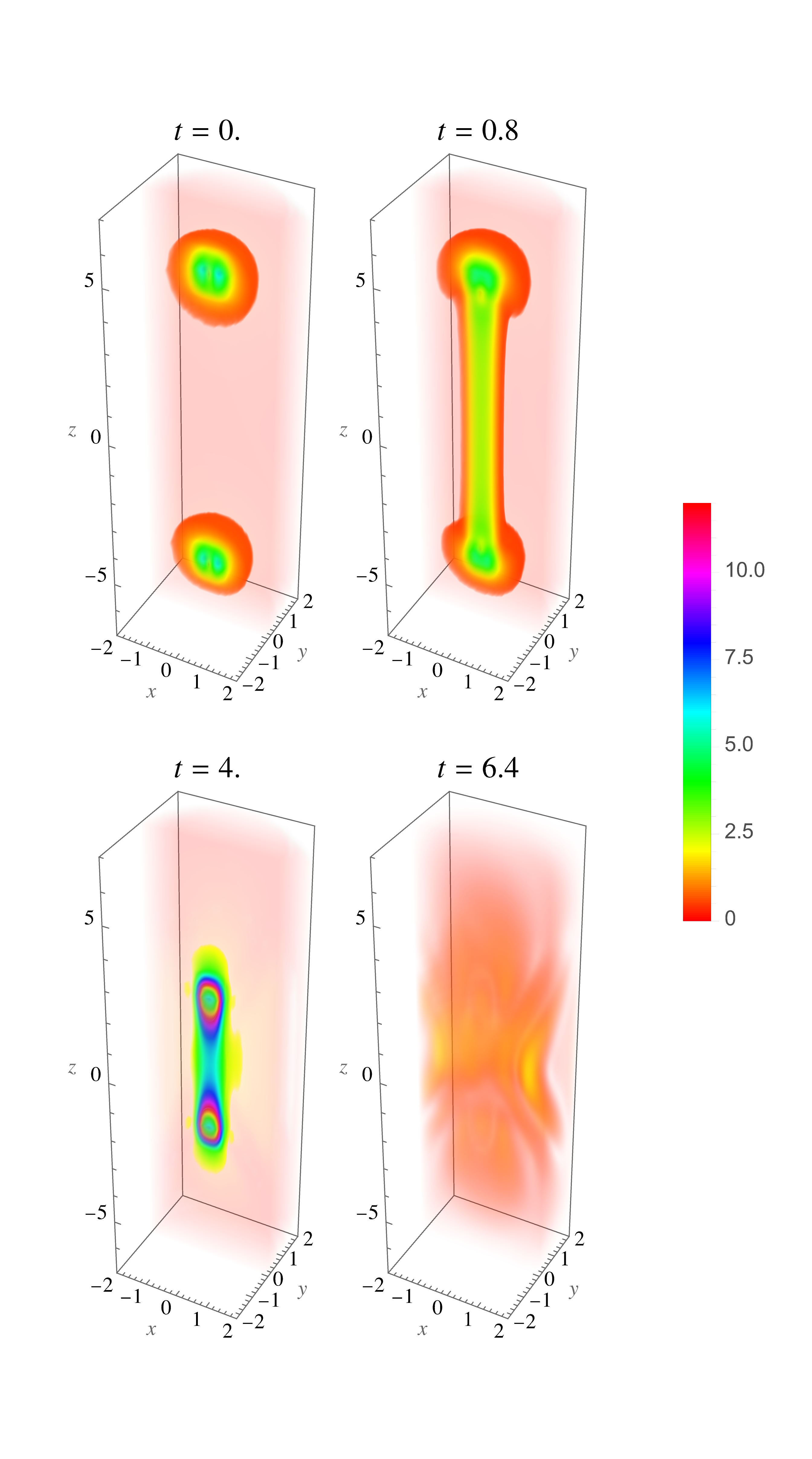}
      \caption{Evolution of the magnetic field norm. The picture is sliced at $y=0$ for better readability. }
    \label{fig:Magnetic Field Density}
\end{figure}
In Fig.~\ref{fig:Magnetic Field Density}, snapshots of the time evolution of the ``magnetic'' field norm are shown. The string formation is observed at early times. For another visual representation of the collapse dynamics see the following~\href{https://www.youtube.com/watch?v=M4IX2JFVpGk}{link} (the sphaleron dynamics described below is also shown there).

Effectively, the system behaves as a type II superconductor and the magnetic field is neutralized by the longitudinal massive photon component which is nothing but the eaten up Goldstone boson. Due to the compactness of the group, as expected, no residual magnetic field is observed. The confining string, being related to the breaking of the $U(1)$ subgroup, carries the same magnetic flux as the original monopoles and is therefore able to fully neutralize the magnetic field.

An example of string formation is shown in Fig.~\ref{fig:winding}, where the initial phase of $\psi$ was randomly chosen at each lattice point. The emergence of the winding further justifies the chosen initial conditions mentioned at the beginning of this section, which is adopted from here on wards. 

\begin{figure}[t]
    \centering
    \includegraphics[width=0.4\textwidth]{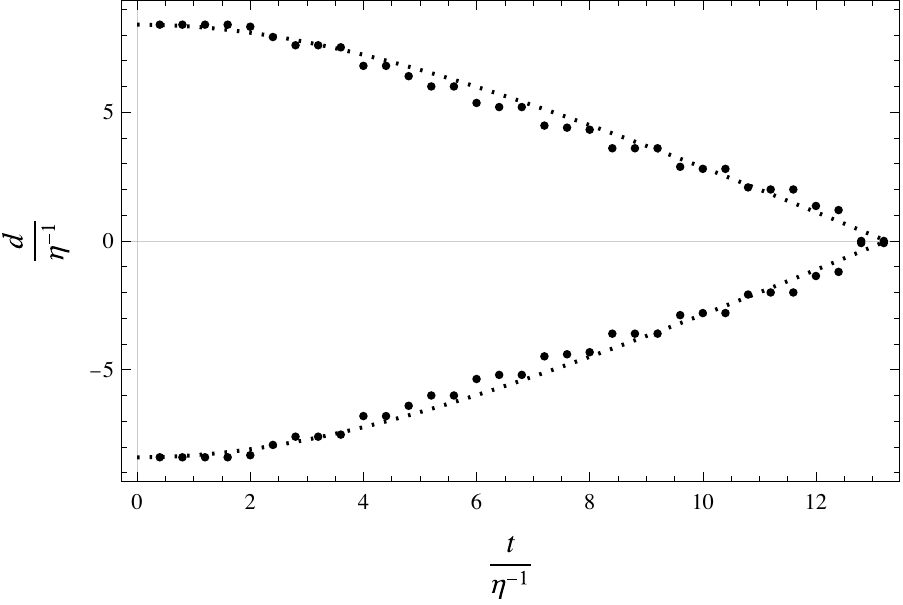}
    \caption{Time evolution of the position of the cores of the monopole and anti-monopole. The dots are numerical results while the dashed line is the point-like solution \eqref{eq:EOM-solution}. Here $a=\mu/m_m=0.16\eta$. }
    \label{fig:position-evolution}
\end{figure}
The dynamics is well approximated by the point-like study of Martin and Vilenkin~\cite{Martin:1996cp}. The system they considered has action
\begin{equation}
\label{eq:pointlike_action}
	I = - m_m \int ds_1 - m_m \int ds_2 - \Lambda^2 \int dS,
\end{equation}
where the first two terms correspond to the monopole/anti-monopole worldline, while the last term describes the string world-sheet for Nambu action\cite{Nambu1977String-likeTheory}. The equation of motion obtained by varying action~\eqref{eq:pointlike_action} admits a particular solution of an exactly straight string with the monopole/anti-monopole pair accelerating towards its center. Their trajectories are given by \cite{Martin:1996cp}:
\begin{equation}
    x(t)=\pm\frac{{\rm{sgn}}(t)}{a}(\gamma_0- \sqrt{1 + (\gamma_0 v_0-a |t|)^2})
    \label{eq:EOM-solution}
\end{equation}
where $t_0= - \gamma_0 v_0/a$,
$a = \Lambda^2/m_m$, $v_0$ and $\gamma_0 = (1-v_0^2)^{-1/2}$ are respectively
the maximum velocity and Lorentz factor of the monopoles, reached at $t = 0$. \\
As shown in Fig.~\ref{fig:position-evolution}, the trajectory followed by the centers of the monopoles (black dots) are nicely fitted by solution~\eqref{eq:EOM-solution} (dashed line).
As expected, the monopoles, due to the constant acceleration, rapidly become relativistic. However, differently from the point-like solution, they annihilate upon reaching approximately zero distance. In contrast, in solution~\eqref{eq:EOM-solution}, they simply  pass through each other and oscillate.


\section{Similarities between confined quarks and 
monopoles} 
Our numerical analysis shows that after the first direct collision,
the monopole and antimonopole always annihilate and decay into waves, without further oscillations.  At first glance, this looks counterintuitive, 
since {\it a priori} there is no reason for monopole and antimonopole  not to pass through each other stretching the string 
of the opposite polarity.

This phenomenon can be explained by the loss of coherence, or equivalently, by the entropy suppression. 
The basic point is that, once the monopole and antimonopole 
come on top of each other, the system looses
coherence due to the emission of waves.
A further re-creation of monopole-antimonopole pair
connected by a string represents a process of a transition from a highly energetic localized source into a coherent state of many soft quanta.  
Very general arguments~\cite{Dvali:2020wqi} indicate that such a process is exponentially suppressed, unless the microstate entropy of the coherent state is close to saturating a certain upper bound set by unitarity. 

In the present case, the microstate degeneracy 
of the coherent state describing the pair of confined monopoles 
is not even close to this value.

Due to this, it is insufficient for matching the phase space occupied by the waves of Higgs and gauge bosons. Correspondingly, rather than creating a highly coherent configuration of monopoles confined by a long string, with much higher probability, the system chooses to decay into those waves. 

Interestingly, the reasoning of~\cite{Dvali:2020wqi} indicates that, in the analogous regime, a similar behaviour 
is expected for the quarks confined by a long QCD string. 
 
In order to display the arguments in the language applicable to both systems, let us consider a $SU(N_c)$ gauge theory with 
a heavy quark transforming in a fundamental representation. 
The term ``heavy" implies  that the mass of the quark, 
$m_q$, is higher than the the QCD scale, $\Lambda$. 
No light quarks are assumed to exist in the theory. The spin of
the quark is not important for our analysis. 

The scale $\Lambda$ sets the boundary 
between the two descriptions.     
At distances shorter than the QCD length, $\Lambda^{-1}$, the viable degrees of freedom are gluons and quarks. At distances larger than $\Lambda^{-1}$, the theory confines and the physical degrees of freedom are colorless composites such as glueballs and mesons. 
 
The phenomenon of confinement is often described as a dual version of the Meissner effect~\cite{Nambu:1974zg,Mandelstam:1974pi,tHooft:1977nqb}.  
The essence of this description is that the gluon ``electric" field gets trapped in a flux tube
when sources are separated by a distance $d \gg \Lambda^{-1}$. 
This is analogous to the behaviour of the magnetic field of monopoles in the $U(1)$-Higgs phase. In the present discussion,  the role of the sources will  be played by a heavy quark-anti-quark pair. 
When separated by a distance $d \gg \Lambda^{-1}$, 
the pair is connected by a QCD string. The thickness of the 
string is set by the QCD length $\Lambda^{-1}$.   The tension of the string is given by the QCD scale, $\mu \sim \Lambda^2$.   
 
Thus, a quark-anti-quark pair connected by a QCD string is strikingly similar to a monopole-anti-monopole 
connected by a magnetic flux tube.
It is a long standing question of how much of QCD physics 
is captured by this analogy. 

In the processes in which the string stays longer than its thickness, the similarity between the two systems is not surprising. 
Naively, this similarity is not expected to extend to the processes in which the string shrinks to the size of its thickness, since physics
governing the two systems at such distances are very different:  
\begin{itemize}
   \item In the case of confined monopoles, the gauge theory is in the Higgs phase. The degrees of freedom are gauge and Higgs bosons. The monopoles, as well as the string connecting them, represent the solitonic objects. 
    \item On the other hand, in QCD the quarks are fundamental particles rather than solitons. Also, the color string connecting them is not describable as a solitonic solution of the classical equations of motion.   
\end{itemize}

Despite these differences, we can find out that the two systems do exhibit similarities even in certain processes controlled by 
physics below and around the scale $\Lambda^{-1}$. 

This similarity concerns the observation in our numerical analysis that in a head-on collision the re-creation of a long string never takes place. This behaviour is expected to be shared by the confined quarks in the similar kinematic regime. 
Below we give supporting arguments. Our discussion will be mostly qualitative.     
In order to keep close to the parameters of our numerical 
analysis, we shall assume that $m_q \sim \Lambda$. Of course, we must assume that quark masses are somewhat larger than $\Lambda$ in order for the long string to be sufficiently stable against the quantum break-up via pair nucleation. 
The system then is characterised by the two scales:
$\Lambda$ and the initial separation of quarks, $d$.
  
A useful  physical way for understanding the suppression is via describing the string-formation process in an effective theory of glueballs and mesons. 
As said, these represent the correct physical degrees of freedom at distances larger than the QCD length. 
In our parameter regime, the characteristic size of an unexcited  glueball or a meson is $\Lambda^{-1}$, and their masses are $\sim \Lambda$. 
 
Qualitatively, one can think of  glueballs and mesons as of closed and open strings of sizes $\Lambda^{-1}$. 
We must however remember that such strings cannot be
described as classical solitons.  For this, it suffices to notice that the Compton wavelengths of glueballs and mesons are $\sim \Lambda^{-1}$. 
We shall assume that 
the initial energy is dominated by the energy of the string, and that the contributions from the quark masses are subdominant. 
That is, 
$d \gg 1/m_q \sim \Lambda^{-1}$.
This implies that in the moment of the collision the quarks are ultra-relativistic. 

The string that is shrunk to the size of its width $\Lambda^{-1}$, represents a blob of energy $\sim d \Lambda^2$. The energy per volume of quark Compton wavelength, is much larger than the masses of quarks.  Correspondingly, quarks are effectively massless. 
Of course, the energy density is also much larger than the masses of unexcited mesons and glueballs. 
Correspondingly, the shrunk string is free to produce mesons and glueballs of high multiplicity, $n \sim d\Lambda$. 

Thus,  system has to decide whether to produce a
single long string or to hadronize into 
$n$ short strings, in form of the glueballs and mesons.  
The outcome is decided by the entropy of the final state.    
This entropy, in case of a long string, is much less than the 
entropy of a generic state with high multiplicity of mesons and glueballs. 

This can be understood from the fact that in an effective 
theory of glueballs and mesons, a long QCD string 
represents a coherent state. 
For $d \gg \Lambda^{-1}$, such a state can
be viewed as approximately-classical.  
The mean occupation number of its constituents can be estimated 
as 
\begin{equation}\label{nglue} 
n \sim d\Lambda\,. 
\end{equation} 
Their characteristic 
de Broglie wavelengths are given by $\sim d$. 
Now, it is intuitively clear that from all possible microstates
of $n$ mesons and glueballs, only a small fraction represents a highly coherent state of a long string.  
Correspondingly, the phase space for string-formation is tiny.  

For better estimate, let us follow the scattering process more closely. 
When the initial string of size $d$ collapses to the size
$\Lambda^{-1}$, it can be viewed as a highly excited 
state of a meson. 
In this language a further formation of a long 
string, represents a transition process from 
a highly energetic quantum meson into a coherent state 
of $n$ soft quanta. According to \cite{Dvali:2020wqi},  
the probability of such a transition process is suppressed by the following universal factor, 
 \begin{equation} \label{crossG} 
\sigma \, \lesssim \, {\rm e}^{-n + S} \,.  
\end{equation}  
The first term ${\rm e}^{-n}$ encodes an exponential suppression 
characteristic of the transitions from a single particle state into
a state of high occupation number $n$. 
The second factor, ${\rm e}^{S}$, comes from the microstate entropy $S$ of the final 
$n$-particle state. Not surprisingly, the entropy enhances the transition probability as it counts the number of the available final states.  

In the present case, the final macrostate we are looking for 
is a QCD string of length $d$. 
For such a string, the
degenerate microstates are given by all possible distinct orientations. Their number  is $\sim (d\Lambda)^2$ and the corresponding microstate entropy is given by,  
 \begin{equation} \label{Sstring}
 S \sim \ln(d^2\Lambda^2) \,. 
 \end{equation}
 Taking into account (\ref{nglue}), we obtain the 
 following estimate for the suppression factor 
   \begin{equation} \label{crossString} 
\sigma \, \lesssim \, {\rm e}^{- d\Lambda + 2\ln(d\Lambda)} \,.   
\end{equation}  
This exponentially small number explains why in head-on collisions the system prefers to hadronize in a high multiplicity of short strings, rather than to produce a single long one.  
  
In other words, due to insufficient entropy, a long QCD string cannot saturate the transition process from a highly energetic meson state. 
As argued in \cite{Dvali:2020wqi}, 
for an unsuppressed production of an $n$-particle coherent state in a transition process of the type, $1\rightarrow n$, 
its microstate entropy must be $S \simeq n$. 
This is clear from  the equation (\ref{crossG}). 
The states with such a high entropy were referred to as ``saturons".
The expression (\ref{Sstring}) shows that the entropy of the long string is much less than the saturation entropy. 
This is the main reason for the suppression. 
Since the production of a long string is exponentially suppressed, the system fills the remaining phase space by a hadronization into a high multiplicity of mesons and glueballs.  

Physics that governs dynamics in the similar kinematic process 
with monopoles is analogous. After the collapse of 
a long string, the system forms a blob with very high energy density. 
It is entropically preferred to dissipate this energy
in high multiplicity of Higgs and gauge bosons rather then in a highly coherent state of monopoles connected by a long string.    

Thus, the suppression of the monopoles-passage through each other, as well as the inability of self-recreation of a 
long QCD string after its fist collapse,  
represent  the phenomena belonging  to the same universality class of effects taking place in high energy collisions.  Their unifying feature  is the exponential suppression of production of coherent states with insufficient microstate entropy~\cite{Dvali:2020wqi}. 

Another phenomenon belonging to the same universality class 
is the inability of a monopole to go through a domain wall,
discussed in~\cite{Dvali:1997sa}. 
In this theory, the domain wall represents a two-dimensional 
sheet in which the Higgs field vanishes and 
the nonabelian (grand unification) symmetry is restored. 
Due to this, when a monopole meets the wall, it is no longer subject to the topological obstruction that  keeps the entire magnetic charge in one point.  Correspondingly, monopole unwinds and the magnetic charge is spread along the wall. \footnote{
Of course, within specific grand unified models,
there can exist monopoles that are not affected topologically by a given domain wall and thereby can pass through without spreading 
the magnetic charge \cite{Brush:2015vda}. These are not relevant for the present discussion.} 
  
In~\cite{Dvali:1997sa} this effect was called  the ``erasure" of topological defects. It was argued that, as a bonus, this mechanism can substantially reduce the cosmological abundance of monopoles
in grand unified theories. In such theories the 
phase transition that forms monopoles, at the same time, 
forms the unstable domain walls.  The walls sweep away monopoles and then disappear. 
A somewhat analogous effect of erasure was observed in $2+1$-dimensions in interactions between skyrmions and domain walls~\cite{Kudryavtsev:1997nw}, and, more recently, between vortices and domain walls~\cite{ourpaper2}.

Although it is both energetically and topologically permitted 
for the monopole (or a skyrmion) to go through the wall and materialize on the other side, the process is highly unlikely. This is also confirmed by more recent numerical analysis of the monopole-wall system~\cite{ourpaper}.
Again, the reason is the loss of coherence and the entropy suppression.  In the example of~\cite{Dvali:1997sa}, the domain wall 
provides a support for unwinding the magnetic charge in the same way as the antimonopole does in the present case.  The 
loss of coherence due to induced waves makes the 
recreation of monopole configuration improbable \footnote{In this respect, it is interesting to compare the situation with scattering of solitons in $1+1$ dimensional theory, which are known to be able to go through each other~\cite{Rajaraman:1982is}.  There are at least two factors that make difference with the monopole case. First is dimensionality which restricts the phase space for the loss of coherence. Unlike monopoles, in case of kinks there is no transverse direction available for emitting the waves. Correspondingly, the precursors that could potentially take away coherence can only travel in the same directions as kinks and cannot escape efficiently. The second factor is confinement.
It will be interesting to study the system within our parameter space.
We thank Tanmay Vachaspasti for raising this question.}.

From the point of view of Ref.~\cite{Dvali:2020wqi}, 
all the considered cases fall in the universal category 
of the processes in which the microstate entropy of the final 
coherent state is not sufficient for compensating the exponential 
suppression~\eqref{crossG}. 
 As discussed, the same conclusion applies 
to a long QCD string viewed as a coherent state in effective theory 
of mesons and glueballs.  

This said, one can certainly imagine the parameter regimes in which the long string can be re-created with less suppression. For example, let us assume $m_q \gg \Lambda$ and the energy of the initial string $=\,3m_q$. In such a case, the quark-antiquark pair moving apart after the first collision, can stretch a string of the length $d \sim m_q/\Lambda^2$. 
This is because the system does not possess the energy 
required for creating an additional quark-antiquark pair, 
necessary for the string breakup. Of course, even in this regime, quarks can annihilate into gluons which then will hadronize, without recreating a long string. However, the annihilation cross-section is perturbative since the scattering length is much shorter than $\Lambda^{-1}$.
 
Also, in case of monopoles, behaviour can be very different in the regime in which the monopole size is much smaller than the width of the string. In this situation, for non-zero impact parameter, the monopole and antimonopole  can miss each other and the string can oscillate, while shrinking to the size of its width. One may expect that the efficiency of annihilation requires that the impact parameter is less than the size of the monopole core. In this case, the annihilation cross-section presumably will be suppressed by a geometric factor\footnote{We thank Alex Vilenkin for commenting.}. However, the understanding of this regime requires a separate study. 

 \section{Gravitational Waves}
 \begin{figure}[t]
    \centering
    \includegraphics[width=0.4\textwidth]{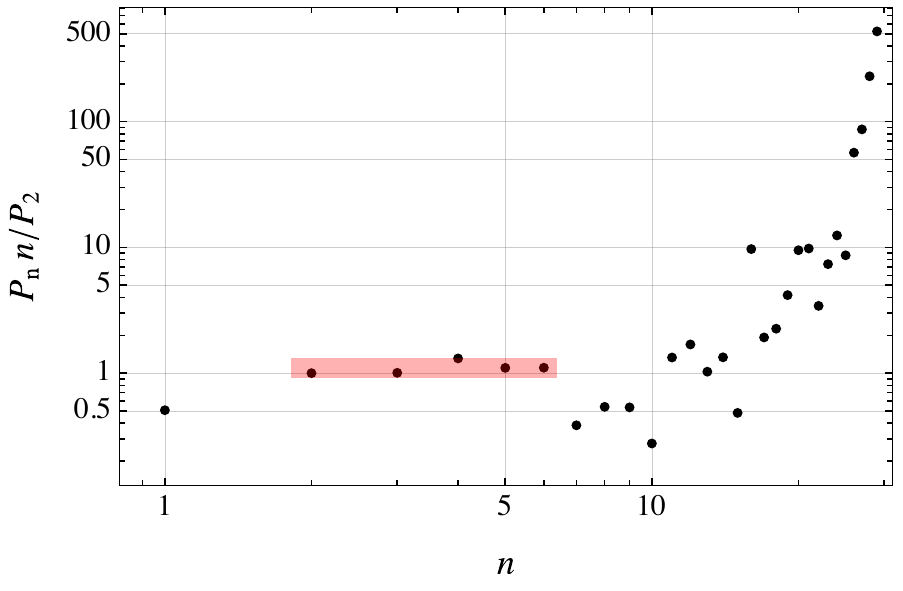}
    \caption{Angularly integrated power emitted $P_{\rm n}$ according to \eqref{eq:radiatedpower}.The amplitude was normalized w.r.t. to the lowest frequency comparable to the initial distance between monopole and antimonopole, up to frequencies comparable to the monopoles size $P_{\rm n}\propto n^{-1}$ (shaded region).}
    \label{fig:GW}
\end{figure}
The radiated power at frequency $\omega_n=2\pi n/T$ ($T$ being the collapse time)\footnote{For practical purposes we chose $T= 2 d \eta^{-1}$, $d$ being the lattice distance between monopoles.}, per unit solid angle in the direction $\textbf{k}$, $|\textbf{k}|=\omega_n$ can be computed as \cite{weinberg1972gravitation}
\begin{equation}
 \begin{split}
     P &= \sum_n P_n = \sum_n \int d\Omega \frac{dP_n}{d\Omega},\\
     \frac{dP_n}{d\Omega}&=\frac{G\,\omega_n^2}{\pi}\left(T_{\mu\nu}^*(\omega_n,\textbf{k})T^{\mu\nu}(\omega_n,\textbf{k}) - \frac{1}{2}|T_\mu^\mu(\omega_n,\textbf{k})|^2\right),
     \label{eq:radiatedpower}
 \end{split}
 \end{equation}
 where the energy-momentum tensor in momentum space is given by
 \begin{equation}
     T^{\mu \nu}(\omega_n,\textbf{k})=\frac{1}{T}\int_0^T dt\, e^{i\omega_n t}\int d^3 \textbf{x} e^{-i \textbf{k}\cdot \textbf{x}}T^{\mu \nu}(t,\textbf{x}).
 \end{equation}
The radiated spectrum obtained from our simulation is shown in Fig.~\ref{fig:GW}, where the dependence of $(P_n\cdot n)$ vs $n$ is shown over logarithmic intervals. We normalised the plot w.r.t $P_2$, corresponding to a wavelength comparable to half of the monopoles initial distance. Direct comparison with the point-like study of Martin and Vilenkin~\cite{Martin:1996cp} shows interesting salient features.

As expected, for low frequencies, $n\lesssim 7$ (red shaded area in the plot), corresponding to an emitted radiation when the monopoles are still far from each other, our result is well approximated by the point-like study. In fact, a spectrum behaviour $P_n \propto n^{-1}$ is observed similar to the point-like case~\cite{Martin:1996cp}.
However, the spectrum is different when the monopoles annihilate. In fact, as shown in Fig.~\ref{fig:GW} the monopole annihilation boosts the radiation spectrum for frequencies comparable to the monopoles and string width. Even though it was not possible to find an exact scaling of the spectrum in this region, this behaviour is different from the point-like case where no enhancement is observed.

The corrections to the spectrum of gravitational waves coming from the finite widths of the string and monopoles, can be of observational interest in cases when this length-scale is macroscopic. For example, a string of the tension 
$\mu = (10^{16}{\rm GeV})^2$ can have a width 
of order km, provided the gauge coupling of the theory 
is $g \sim 10^{-35}$. Existence of such super-weak gauge interaction  in 
some hidden sector of the theory is fully consistent with all known laws of physics. Of course, the parameters in our simulation could not be taken with such extreme values. However, the dynamic of the string can be significantly affected due to the difference of Higgs and magnetic cores  as in the regimes discussed in~\cite{Adelberger:2003qx}.
Correspondingly, applications 
of our results for such parameter ranges, can only be viewed as indicative. 


\section{Sphaleron}
\begin{figure}[t]
    \centering
    \includegraphics[width=0.45\textwidth]{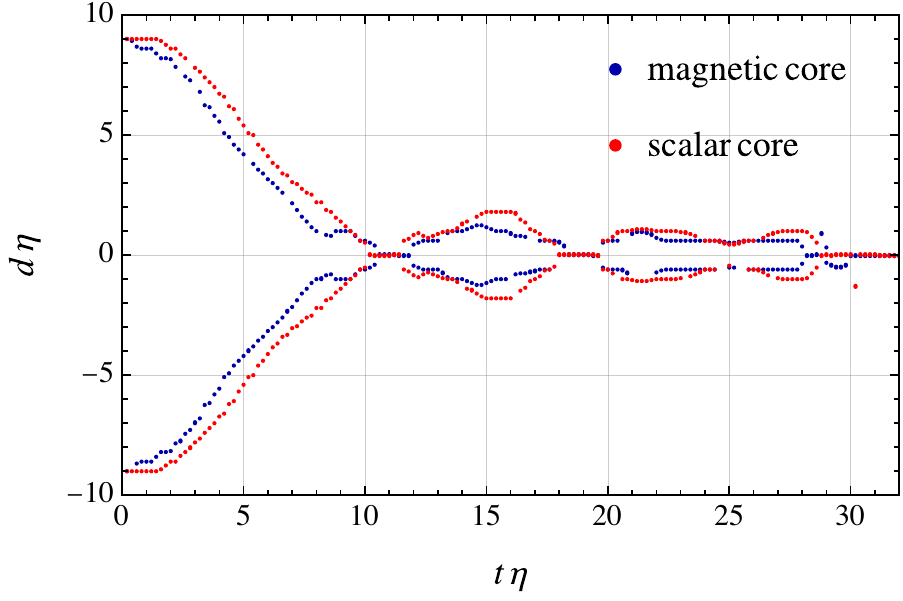}
    \caption{Evolution of the magnetic and scalar cores of the monopole/anti-monopole pair for initial twist $\gamma = \pi$.}
    \label{fig:sphaleron}
\end{figure}
So far we have been focused on the untwisted configuration i.e., $\gamma = 0$. This corresponds to a minimum of the string energy. While it is obvious from \eqref{eq:String-Ansatz2} that rotating $\gamma$ by $2\pi$ gives back the same configuration, it is straightforward to see how the energy varies with $\gamma$ (at fixed distance $d$) and find that $\gamma = \pi$ is an energy maximum.

The role of the twist $\gamma$ has a deep topological meaning. There exists a static bound state configuration known as ``sphaleron" -- see, eg.,~\cite{Manton:2004tk} for a review and~\cite{Klinkhamer:1984di,Manton:1983nd,Brihaye:1994ib} for solutions in the electroweak sector. This bounded configuration interpolates between vacua of different Chern-Simons number. It corresponds to the maximum of the non-contractible loop interpolating between two such minima and has Chern-Simons number $1/2$~\cite{Manton:1983nd}. 

It was argued in~\cite{Taubes:1982ie} that such unstable solutions can be understood as a monopole and a rotated antimonopole. For the maximal twist $\gamma = \pi$, the Chern-Simons number of \eqref{eq:MMbar-Ansatz-W} corresponds to that of a sphaleron as it can be easily checked. 

Refs.~\cite{Saurabh2017Monopole-antimonopolePotential,Kleihaus:1999sx} have numerically verified the behaviour of this object in the unconfined case. They observed that the twist
 $\gamma=\pi$ prevents the monopole-antimonopole annihilation. To our knowledge, the behaviour of this configuration in the confined phase has not been studied so far. 

In Fig.~\ref{fig:sphaleron}, the position of the two monopoles core centers is shown as a function of time. As it can be seen, at initial time the two magnetic cores start to accelerate towards each other due to the flux tubes connecting them, dragging the scalar core along and the system becomes relativistic. 

The dynamics is analogous to the untwisted one, c.f. Fig.~\ref{fig:position-evolution}, up to the collapse moment. While in the untwisted case the two monopoles annihilate right away, in the twisted case they do not. Instead, the magnetic cores repel each other, slow down and bounce back. 

From there on the two monopoles have a bouncing behaviour, and dissipate energy up to settling to a constant distance where the confining energy is balanced by the repulsive twist energy. Eventually, the pair fully annihilates. Concomitantly a deviation from axial symmetry is observed. Given the axial symmetry of the initial configuration, and the fact that the dynamics should preserve such symmetry, we believe this to be a numerical artefact.

In Fig.~\ref{fig:sphaleron}, $\Lambda\simeq\eta$ was used. For lower value of this parameter the bouncing sizes, as well as distance of the bound state were observed to be larger. For a different initial distance, 2-dimensional slices of the dynamics can be found at the following~\href{https://www.youtube.com/watch?v=M4IX2JFVpGk}{link}.

For $\gamma\neq\pi$, no bounce is observed, as the system can always untwist in the direction of favourable minimum, thereby annihilating right away.

\section{Conclusion and outlook}
In this article we numerically studied the confinement of a $SU(2)$ monopole/anti-monopole pair. Initially the pair is in a dipolar configuration as in Fig.~\ref{fig:examplephi}. The confinement is instantaneously imposed by setting an extra complex doublet field in its Higgs-phase, therefore breaking the residual $U(1)$ and giving mass to the previously massless ``photon".

As shown in Fig.~\ref{fig:position-evolution}, 
very quickly the monopoles become relativistic due to confinement. Taking initial random phases for the confining scalar field, it is possible to see the dynamical emergence of the tube as shown in Fig.~\ref{fig:winding}.

The GW spectrum is also computed and is compared with the point-like study of Martin and Vilenkin~\cite{Martin:1996cp}. Before the collision and for the wavelengths longer than the relevant widths of the system (i.e., monopoles and string width) we find perfect agreement. Namely, we confirm a power spectrum $P_n \propto n^{-1}$, $n$ denoting the frequency number. As previously mentioned, this type of spectrum can lead to a flat density of GWs $\Omega_{GW}$ across several orders of the frequency, therefore making it interesting from a phenomenological perspective in view of future refinement of the stochastic GW background hints obtained from pulsar timing arrays~\cite{NANOGrav:2020bcs,Antoniadis:2022pcn}. 

However, in our parameter regime, in which the scales of string and monopoles are of the same order, we observe the following differences  from the point-like case. 

Firstly, an enhancement of the GW spectrum for wavelengths comparable to the monopoles and string size as shown in Fig.~\ref{fig:GW}, therefore providing finite-width corrections to the point-like case~\cite{Martin:1996cp}.
This can be of phenomenological interest for the strings of a macroscopic width.  Notice that such strings can have a high tension provided the gauge coupling is sufficiently weak and can produce 
intense gravitational waves.

Secondly, in our numerical analysis we have observed an interesting effect. 
After the first collision, the monopoles annihilate    
without any further oscillations. 
We gave a qualitative explanation 
to this observation in terms of a general phenomenon 
of suppressed production of low-entropy coherent states  
in a collision process~\cite{Dvali:2020wqi}. 
In the present case, the loss of coherence in the monopole-anti-monopole collision,  makes the recreation of a monopole-string system unlikely due to an insufficient entropy of such a state. 
 
Applying the general reasoning of~\cite{Dvali:2020wqi}, we argued that in the similar parameter regime this behaviour must be shared by heavy quarks confined by a long QCD string. 
There too, after the collapse of a straight long string, instead of 
re-stretching it in an oscillatory mode, the system prefers to 
directly decay into high multiplicity of mesons and glueballs.  
This similarity extends the connection between the confined quarks and monopoles to the domain of processes controlled by the short-distance physics.

\textit{\textbf{Aknowledgements}}: We thank Tanmay Vachaspati and Alex Vilenkin for useful discussions and valuable comments.\\
This work was supported in part by the Humboldt Foundation under Humboldt Professorship Award, by the
Deutsche Forschungsgemeinschaft (DFG, German Research Foundation) under Germany’s Excellence Strategy - EXC-2111 - 390814868, and Germany’s Excellence
Strategy under Excellence Cluster Origins.
\bibliography{references}
\begin{widetext}
\appendix
\section{Fields equation}\label{app:Fields_eq}
\subsection{Field equations}
The field equations can be written as
\begin{align}
    \partial_t^2\varphi^a &= 
    \nabla ^2\varphi^a
    -g\epsilon^{abc}\partial_i \phi^b W_i^c 
    -g \epsilon^{abc} (D_i\varphi)^b W_i^c
    -\lambda(\varphi^b\varphi^b-\eta^2)\varphi^a
    -g \epsilon^{abc}\varphi^b\Gamma^c
    -c\psi^\dag \sigma^a\psi,
    \label{eq:EOM-Varphi}\\
    \partial_t^2\psi^\alpha &= 
    \nabla^2 \psi^\alpha
    -\frac{g^2}{4} W_i^a W_i^a \psi^\alpha 
    - i\frac{g}{2} \Gamma^a (\sigma^a\psi)^\alpha 
    - i g W^a_i (\sigma^a\partial_i\psi)^\alpha
    - c \varphi^a \sigma^a \psi^\alpha
    -\tilde{\lambda}(\psi^\dag\psi-v^2)\psi^\alpha,
    \label{eq:EOM-Psi}\\
    \partial_t W^a_{0i}&=
    \nabla^2 W_i^a 
    + g \epsilon^{abc} W_j^b \partial_j W_i^c -g\epsilon^{abc} W^b_j W^c_{ij} 
    -D_i \Gamma^a 
    - g \epsilon^{abc}\varphi^b(D_i\varphi)^c
    -\frac{1}{2}g^2\psi^\dag\psi W_i^a-i g \psi^\dag\sigma^a\partial_i\psi,
    \label{eq:EOM-W}\\
    \partial_t \Gamma^a &=
    \partial_i W_{0i}^a
     -g_p^2\left[\partial_i(W_{0i}^a)+g\epsilon^{abc} W^b_i W^c_{0i}+ g \epsilon^{abc}\varphi^b(D_t\varphi)^c+i g\psi^\dag\sigma^a \partial_t \psi\right],
    \label{eq:EOM-Gamma}
\end{align}
where we are using the temporal gauge, $W^a_0 = 0$, $\Gamma^a = \partial_i W^a_i$ are introduced as new variables, and $g^2_p=1.5$ is a numerical parameter that we can choose to ensure numerical stability. The equations were evolved with a Cranck-Nicolson leapfrog algorithm combined with absorbing boundary conditions \cite{ISRAELI1981115}.
\begin{figure}[h]
     \centering
         \includegraphics[width=0.8\textwidth]{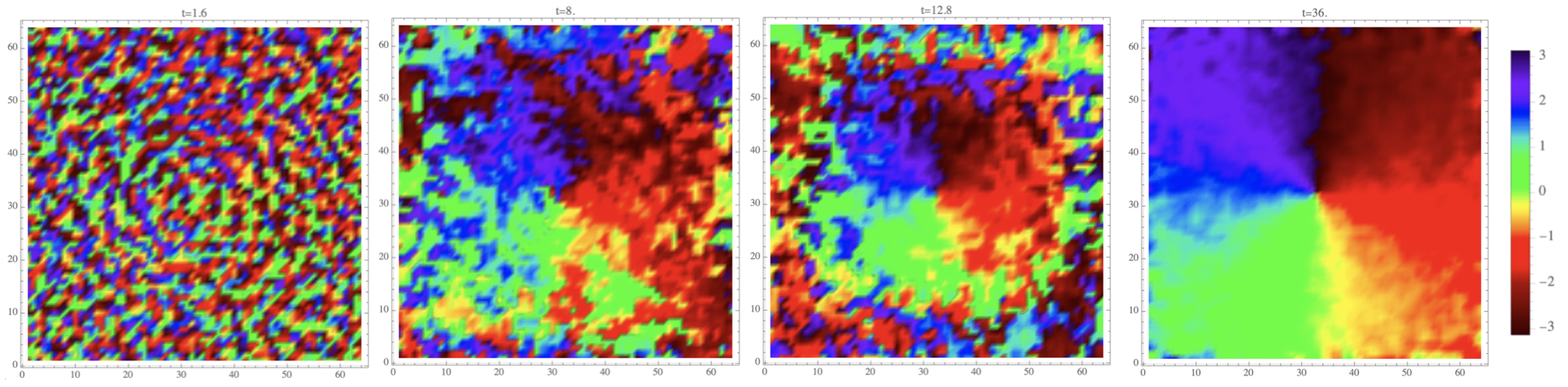}
      \caption{Phase of the confining scalar field in the \textit{xy} plane at the middle of the string height. }
    \label{fig:winding}
\end{figure}
\end{widetext}
\end{document}